# AN OBJECTIVE CLASSIFICATION OF SATURN CLOUD FEATURES FROM CASSINI ISS IMAGES

Anthony D. Del Genio[a*] and John M. Barbara[b]

[a]*NASA Goddard Institute for Space Studies, New York, NY 10025 USA*

[b]*Trinnovim LLC, Institute for Space Studies, New York, NY 10025 USA*

[*]Corresponding author: NASA Goddard Institute for Space Studies, 2880 Broadway, New York, NY 10025 USA. Phone: 212-678-5588.  Email: anthony.d.delgenio@nasa.gov

ABSTRACT

A k-means clustering algorithm is applied to Cassini Imaging Science Subsystem continuum and methane band images of Saturn's northern hemisphere to objectively classify regional albedo features and aid in their dynamical interpretation.  The procedure is based on a technique applied previously to visible-infrared images of Earth. It provides a new perspective on giant planet cloud morphology and its relationship to the dynamics and a meteorological context for the analysis of other types of simultaneous Saturn observations.  The method identifies 6 clusters that exhibit distinct morphology, vertical structure, and preferred latitudes of occurrence.  These correspond to areas dominated by deep convective cells; low contrast areas, some including thinner and thicker clouds possibly associated with baroclinic instability; regions with possible isolated thin cirrus clouds; darker areas due to thinner low level clouds or clearer skies due to downwelling, or due to absorbing particles; and fields of relatively shallow cumulus clouds. The spatial associations among these cloud types suggest that dynamically, there are three distinct types of latitude bands on Saturn: deep convectively disturbed latitudes in cyclonic shear regions poleward of the eastward jets; convectively suppressed regions near and surrounding the westward jets; and baroclinically unstable latitudes near eastward jet cores and in the anti-cyclonic regions equatorward of them. These are roughly analogous to some of the features of Earth's tropics, subtropics, and midlatitudes, respectively. This classification may be more useful for dynamics purposes than the traditional belt-zone partitioning. Temporal variations of feature contrast and cluster occurrence suggest that the upper tropospheric haze in the northern hemisphere may have thickened by 2014. The results suggest that routine use of clustering may be a worthwhile complement to many different types of planetary atmospheric data analysis.

Highlights:

- K-means clustering is used to classify all Saturn image areas meteorologically

- Six cloud types similar to those in terrestrial tropics and midlatitudes are identified

- Cloud type occurrences suggest three dynamically distinct types of latitude bands

# 1. INTRODUCTION

Knowledge about the dynamics and clouds of planetary atmospheres has evolved very differently for Earth compared to other planets in the solar system. Understanding of Earth's general circulation and its relationship to clouds was built on *in situ* surface meteorological observations, upper air soundings, and surface observations of sky conditions (Lorenz, 1991). Only in the past few decades did satellite observations come into play to provide a global context for theories that had already been developed from the *in situ* data.

For other planets, and particularly the giant planets, *in situ* observations are recent, scarce or completely absent. Much of our understanding of the tropospheric dynamics of these planets has therefore relied on satellite imaging. As a result, this knowledge has often depended on subjective interpretations of cloud morphological features, sometimes but not always aided by quantitative information on their temporal evolution or appearance at different wavelengths. Because this approach is heavily influenced by human perception, dynamical inferences from image analysis often focus on spectacular features – Jupiter's Great Red Spot, white ovals, barges, hot spots, and equatorial plumes (e.g., see the review by Vasavada and Showman, 2005); Saturn's polar hexagon, south polar vortex, ribbon wave, string of pearls, and Great Storms (e.g., see the review by Del Genio et al., 2009); and so on.

While this strategy has been very productive, it has two disadvantages: (1) It promotes a somewhat balkanized view of a planet's atmosphere, with each regional feature being analyzed in isolation; (2) it emphasizes unusual, obvious features at the expense of the potential information to be obtained from the smaller, lower contrast but more numerous features present in any image. A more objective approach to planetary atmospheric image analysis may prove to be a useful resource for more fully exploiting the meteorological information content in planetary images and contributing to a more unified, global view of a giant planet's dynamics.

One such objective approach is k-means clustering (Anderberg, 1973). Given an array of image pixel reflectance values (e.g., in spectral space), a k-means algorithm finds an optimal set of clusters, each containing individual pixels or groups of pixels with similar multi-spectral characteristics, that are most common in the data. In principle the clusters can provide insight into cloud properties and the physical processes that produce them and identify similarities between cloud features in very different regions that may otherwise not be apparent.

The first planetary application of k-means clustering was by Thompson (1990), who used 0.5° resolution Voyager 2 Jupiter images in 4 visible filters to define clusters of pixels with similar color/albedo properties. He found 25 distinct color units associated with well known but also more subtle morphological features. Banfield et al. (1998) did not perform cluster analysis, but defined deep cloud features in scatter plots and false color images of weak methane band and continuum reflectances. This technique has been used to identify features of interest on Jupiter and Saturn, e.g., deep convective storms (e.g., Gierasch et al., 2000; Porco et al., 2003; Baines et al., 2009; Sayanagi et al., 2013). Dyudina et al. (2001), Simon-Miller et al. (2001), and Irwin and Dyudina (2002) used empirical orthogonal analysis to categorize Jupiter infrared spectra and infer cloud colors and structure over the disk. These approaches complement traditional radiative transfer studies of global cloud optical properties and top pressures, e.g., Banfield et al. (1998) for Jupiter, and Pérez-Hoyos and Sánchez-Lavega (2006) and Roman et al. (2013) for Saturn. Cluster analysis offers a different objective way to classify both obvious features and others that are less visually prominent.

Objective techniques can also be applied to infer local dynamical processes responsible for cloud albedo contrasts. For Earth, Rossow et al. (2005) used k-means clustering to create a climatology of "weather states." Using visible and window infrared channel weather satellite images, they applied clustering to 2-dimensional frequency histograms of pixel optical thickness and cloud top pressure retrieved from these channels, aggregated over 2.5°x2.5° latitude-longitude areas. The k-means clustering algorithm finds an optimal number of commonly occurring patterns in these histograms. Figure 1 shows the 6 weather state clusters for Earth's tropics (Chen and Del Genio, 2009), while Figure 2 shows maps of the relative frequency of occurrence (RFO) of each cluster and Figure 3 a schematic cross-section of the tropics (Boucher et al., 2013) that interprets the clusters in terms of specific cloud types.

[Fig. 1 approximately here.]

[Fig. 2 approximately here.]

[Fig. 3 approximately here.]

Since the clusters are composed of many pixels aggregated over a mesoscale area, they contain multiple cloud types (hence the broad distribution of optical thicknesses and cloud top pressures in each cluster in Figure 1), but each one can be characterized by its prevalent cloud type. Clusters C1 and C2 are dominated by high top, optically thick and moderately thick clouds, respectively, that are concentrated in the Intertropical Convergence Zone (ITCZ). These are the deep convective cloud systems and associated anvils and cirrostratus clouds that form the rising branch of the Hadley cell. C3 also occurs in the ITCZ but more often on the edges than C1 and C2 and more often over the continents, where humidity is somewhat lower. These clouds have somewhat lower tops and optical thicknesses; they are isolated congestus and moderate deep convective cells that characterize Hadley cell inflow and other transitional regions. C4 are very high, optically thin cirrus that are associated with nearby strongly convecting regions but only where the upper troposphere is sufficiently humid. C5 and C6 are low-topped clouds that occur in the subsiding branches of the Hadley and Walker cells. C5 (frequent occurrence, small cloud cover) are fair weather shallow cumulus that populate the open subtropical oceans, while C6 (infrequent occurrence but large cloud cover) are the marine stratocumulus that dominate the cooler eastern oceans.  Similar cluster analyses have been performed for a more extended region of Earth's subtropics and for the extratropics, adding several more distinct low cloud weather states as well as states associated with baroclinic frontal systems (Williams and Tselioudis, 2007; Oreopoulos and Rossow, 2011).

The success of k-means clustering in producing a simple but objective classification of cloud morphological properties associated with well-known dynamical conditions on Earth suggests that it may be useful to apply to the giant planets, for which we do not yet have the same depth of understanding as embodied in Figure 3. In this paper we apply the Rossow et al. (2005) k-means clustering algorithm to Cassini Imaging Science Subsystem (ISS) images (Porco et al., 2004) to define Saturn weather states and consider similarities and differences with respect to Earth.  Section 2 describes the dataset and analysis method. Section 3 applies the method to simultaneous continuum and methane band images. Section 4 discusses several variations on the results, and Section 5 summarizes the conclusions that emerge from our analysis.

## 2. DATA AND METHODS

### 2.1 Image selection and processing

Unlike for Earth, where high resolution imagery exists in both the visible and thermal infrared, Cassini does not obtain thermal infrared maps at resolutions comparable to those possible with the ISS near apoapsis. We therefore use near-infrared continuum (CB2, 750 nm) ISS images to provide reflectances that are sensitive to cloud optical depth, and simultaneous methane band (MT2, 727 nm) images for which the strength of gaseous absorption causes reflectance to be sensitive to cloud top pressure, as shown in Figure 13 of Roman et al. (2013). In the absence of clouds, CB2 sees to great depths on Saturn and thus senses the first optically thick cloud deck, albeit with some obscuration by the overlying hazes, while MT2 senses only down to the upper troposphere, in the midst of Saturn's upper tropospheric haze (e.g., Fig. 7.9 of West et al., 2009). Therefore, in locations in which clouds have limited vertical extent, CB2 and MT2 sense independent meteorological phenomena at different altitudes, but when clouds and their vertical motions extend over great depth, both filters observe the same dynamical process. Absorbing materials may also affect CB2 reflectance, e.g., in the neighborhood of deep convection (Baines et al., 2009).

We analyze a total of 102 images in each filter from the ISS Narrow Angle Camera (NAC) and Wide Angle Camera (WAC) obtained on 6 November 2007, 17 April 2008, 7 May 2008, and 16 September 2014 at resolutions ranging from 17-89 km and phase angles from 20°-46° (Table 1). These dates were chosen to allow us to focus analysis on the northern hemisphere, which had a thinner upper tropospheric haze and thus greater feature contrast than the southern hemisphere had through most of the seasonal cycle observed by Cassini (see Fig. 1 in Del Genio and Barbara, 2012), thereby allowing for greater altitude discrimination between CB2 and MT2. Our analysis spans planetocentric latitudes 18°-60°N, limited by ring shadow on the equatorial side and decreased illumination on the poleward side.

[Table 1 approximately here.]

Image calibration, photometric correction and navigation are described fully in Del Genio et al. (2007). We summarize the main features here. Cassini Imaging Science Subsystem Calibration software (West et al., 2010) is used to subtract dark current and bias, divide by a flat field, correct for nonlinearity and dust rings, adjust to absolute calibrations when available, and convert to I/F using the insolation at Saturn's distance as a reference. A Minnaert function is applied to partly correct for large-scale illumination gradients. Navigation of NACs exploits near-simultaneous WACs to locate the limb with sub-pixel precision. We match approximate predicted pointing from binary C-kernels to the limb curve with a least-squares fit, then apply a small WAC-NAC boresight correction. Navigated images are then mapped into a cylindrical (rectangular) projection in the planetocentric latitude system.

To remove the remaining large-scale illumination gradient, we subtract a smoothed image (boxcar average with width = 300 pixels) and then add back the mean I/F to eliminate negative brightness values. The resulting images still retain large-scale belt-zone brightness contrasts, especially in MT2. These contrasts are of interest in their own right for understanding cloud/haze composition and particle properties, but they overwhelm the small-scale feature contrasts that contain the dynamical information we seek. We therefore subtract the zonal mean brightness at each latitude and then again add back the global mean I/F to produce a scene with the small-scale features that are our focus but with otherwise uniform brightness on larger scales.

The resulting images thus capture relative variations in brightness on small scales but do not retain the original I/F information needed to constrain cloud radiative properties. Consequently our analysis is concerned only with whether clouds are thicker or thinner, and their tops higher or lower, than nearby clouds in the same scene, and what this suggests about their relationship to the local dynamics. It is not used to retrieve absolute values of optical thickness, cloud top pressure, cloud color, or large-scale cloud differences between belts and zones. Although our approach is patterned after the terrestrial study of Rossow et al. (2005), which does use retrievals of optical thickness and cloud top pressure (Rossow and Schiffer, 1991), earlier analyses of the same terrestrial image data used the original two-channel visible reflectances and infrared radiances themselves instead, e.g., to study convective clouds (Fu et al., 1990).

After the steps described above, the brightness distributions for 2007, 2008, and 2014 still exhibit some differences. The 2007 images are somewhat brighter than the 2008 images and the 2014 images brighter than both other periods, especially in CB2 (Fig. 4). We therefore shifted the 2008 and 2014 brightness distributions by an amount equal to the difference between their mode values and that of the 2007 images. The 2008 CB2 distributions are also broader than those from 2007 and the 2014 CB2 distributions narrower, despite an identical gain state and identical or similar phase angles (Table 1). The 2007 images are NACs and the 2008 and 2014 images WACs, which cannot explain the different behavior of 2008 vs. 2014. Exposure durations were longer for the NACs, but for both the CB2 and MT2 images, yet only CB2 exhibits a noticeable difference in distribution width. We have chosen not to apply a correction for this; it is possible, for example, that this difference is physical, associated with changes at the CB2 level or in the overlying haze as the season progresses. We will return to this topic in Section 3.3.

[Fig. 4 approximately here.]

*2.2 K-means clustering algorithm*

We first divide the 2-filter NAC mosaics and WAC images into 1°x1° latitude-longitude boxes and divide the full range of brightness values in the images into 8 CB2 bins and 7 MT2 bins. We then create 8x7 frequency histograms of the CB2 and MT2 brightnesses of the pixels in each box. The box size is chosen to be just large enough to encompass individual morphological features of interest but small enough to retain minimal overlap in CB2-MT2 brightness value pairs. Saturn has much lower feature contrast than Earth, or even Jupiter, because of its optically thick upper tropospheric haze, and thus the brightness distribution is concentrated near intermediate values, as can be seen in Figure 4. We therefore define the bin widths for the 8x7 histograms so that all CB2 bins contain a roughly equal number of pixels, and likewise for the MT2 bins. The bins are thus unevenly spaced in brightness. The 8x7 dimension of the histogram was chosen subjectively after tests of several other sizes to best isolate areas with extreme low and high brightness values in the two filters. These histograms are the input to the k-means algorithm.

The k-means analysis is initiated by randomly selecting a predetermined number of initial centroids (k) in the CB2-MT2 brightness space. A test with a different set of randomly calculated initial centroids produced clusters whose correlation coefficients with those shown exceeded 0.999. Euclidean distances to each centroid are calculated for each image histogram and the histogram is then assigned to the closest centroid, i.e., the centroid whose pattern in CB2-MT2 brightness histogram space most resembles it. In this way a cluster of points (each point representing one 1°x1° box) surrounding each centroid is built up over all available images. The centroids are then re-calculated for each cluster, the points are re-assigned to the new centroids, and the procedure is iterated until the iteration converges, defined as when distances between two successive generations of centroids fall below 0.001. The procedure ensures that individual cluster members are closer to the centroid of their cluster than any two centroids are to each other, although pixels of similar CB2 and MT2 brightness in different locations may be assigned to different clusters depending on the small-scale feature contrast of neighboring pixels.

### 3. SATURN CLUSTER PROPERTIES

*3.1 Cluster number selection*

The k-means procedure has one subjective element – the user must define the number of clusters k at the outset.  The bigger the dataset, the more clusters one might expect to find as enough "extreme events" occur for them to constitute their own weather states.  To introduce some objectivity into this step, we follow Rossow et al. (2005) and perform the cluster analysis multiple times, for increasing values of k.  For each k, we search for pattern correlations between each pair of cluster histograms that become statistically significant to decide when to stop adding new clusters.

To illustrate, Figures 5 and 6 show the Saturn clusters (designated SC1, SC2, etc. to differentiate them from the Earth clusters in Figure 1) produced when k = 5 and k = 6, respectively. Table 2 shows the corresponding correlations between clusters.  For the 8x7 arrays we use, a correlation of 0.26 (0.34) is significant at the 95%(99%) level.  According to Table 2, clusters SC3 and SC4 are correlated at close to 95% significance. Visual inspection of Figure 5, though, shows that these are mostly independent populations of pixels, both defined by being dark in CB2 but with one primarily bright in MT2 and the other primarily dark in MT2.  The non-negligible correlation comes from pixels of intermediate MT2 brightness in both histograms.

[Fig. 5 approximately here.]

[Fig. 6 approximately here.]

[Table 2 approximately here.]

Comparing Figures 5 and 6, it is obvious that 4 of the 5 clusters for the k = 5 case are retained in very similar form, and with very similar RFO, for k = 6. This indicates that the clustering method is stable for the sample size we have and is selecting physically meaningful repeating CB2-MT2 relationships. The effect of adding a sixth cluster is that cluster SC2 for the k = 5 case, which collects many of the pixels of intermediate CB2 into one cluster regardless of MT2 brightness, is now split into two clusters, one weighted more toward brighter MT2 (SC2) and the other weighted toward intermediate MT2 (SC3). This seems to be a meaningful separation. In fact the correlation between the two new clusters is minimal (-0.06). Two pairs of clusters have marginally significant correlations: SC1 and SC6, which are both bright in CB2 but largely concentrated at opposite ends of the MT2 range; and SC4 and SC5, which are both dark in CB2 but concentrated at higher vs. lower MT2 brightnesses. We also calculated a cluster set for k = 7 (not shown). This set retains the general characteristics of the clusters in the k = 6 case but adds a $7^{th}$ cluster defined by low MT2 and intermediate CB2 I/F. This set contains 3 cluster pairs with correlations 0.36, 0.34, and 0.27, so k = 6 is probably optimal for the size of our dataset. Selecting k = 6 also has the advantage of providing a cluster set that is somewhat analogous to that for the terrestrial data (Fig. 1), which facilitates thinking about similarities and differences between Saturn and Earth.

*3.2 Interpreting the clusters*

From this point forward we focus on the k = 6 clusters. We have numbered the clusters to be consistent with the numbering convention of the Earth clusters in Figure 1. This indicates only similarity in radiative properties (e.g., C1 and SC1 contain the most optically thick, high top clouds for both Earth and Saturn); however, it facilitates our consideration below of the extent to which cloud radiative properties are diagnostic of dynamical mechanisms of formation.

This question has already been posed for Earth, where independent meteorological information has allowed the utility of cluster classification to be evaluated against observed temperature profiles and column water vapor (Jakob et al., 2005), precipitation rates (Jakob and Schumacher, 2008; Tromeur and Rossow, 2010; Lee et al., 2013), cloud radiative forcing (Oreopoulos and Rossow, 2011), and diabatic heating and moistening/drying profiles (Stachnik et al., 2013). General impressions from these studies, as well as earlier comparisons of the histograms (without clustering) to human surface-based cloud observations (Hahn et al., 2001), are that (a) the classification successfully distinguishes meteorologically "disturbed" convective conditions (C1, C2, C3) from "suppressed" conditions dominated by thin cirrus (C4) or low clouds (C5, C6); (b) the most extreme, heavily precipitating organized convective storms are captured by a single cluster (C1); (c) although shallow fair weather scattered cumulus regimes can successfully be separated from near-overcast stratocumulus regimes climatologically (e.g., the bottom two panels of Figure 2), their optical thickness distributions sufficiently overlap that optical thickness alone cannot distinguish them on an instantaneous basis.

For Saturn such ancillary information is sparser. What we do have are visual impressions of cloud morphology that can rule in or out certain cloud types, information on wind velocities and shear at cloud level, and temperatures and composition near and above cloud level (e.g., see the review by Del Genio et al., 2009). We also have spatial associations (mostly latitudinal) between individual clusters. To aid our interpretation, Figures 7 and 8 show one of the WAC CB2 and MT2 images used in the clustering analysis. Superimposed on these images are boxes indicating locations of identifications of each of the 6 Saturn clusters.

[Fig. 7 approximately here.]

[Fig. 8 approximately here.]

Figure 9 distills this information by showing the latitudinal dependence of the number of occurrences of each cluster, along with the Saturn CB2 mean zonal wind profile derived by applying the cloud tracked wind algorithm described in Del Genio and Barbara (2012) to the 1°x1° boxes used for the clustering.

[Fig. 9 approximately here.]

Just as on Earth, a mix of cloud types is found at any latitude. We therefore use relative occurrences of different clusters at different latitudes as a guide to the meteorological context for each cluster in two different ways. First, Table 3 presents the correlation coefficients between the number of occurrences of each cluster at a given latitude (see Fig. 9) and those of all the other clusters. Second, we divide the analysis domain into three different types of latitude bands: Cyclonic shear regions (18-31°N, 44-47°N, 58-60°N), eastward jet cores + anti-cyclonic shear regions (38-44°N, 52-58°N), and westward jet regions (31-38°N, 47-52°N). For each type of latitude band, Table 4 shows ratios of the number of occurrences of each cluster per degree latitude to the analogous number of occurrences for the entire domain. With this information, we consider each Saturn cluster in turn below:

[Table 3 approximately here.]

[Table 4 approximately here.]

*SC1:* This cluster consists primarily of clouds that are bright in both filters and are thus optically thick with high cloud tops. Figure 7 shows that their morphology is mostly isolated cells or compact groups of cells in CB2, many of them visible at reduced levels with the same morphology in MT2. SC1 occurs less frequently (RFO = 6.1%) than any other cluster (Figs. 6, 9), and most often at latitudes with cyclonic shear (Fig. 9, Table 4). All of this is consistent with the interpretation that SC1 locations are primarily deep convective clouds that extend over many hundreds of millibars if not considerably more and may produce significant precipitation and latent heating. SC1 clouds are analogous to those frequently seen in cyclonic shear regions on Jupiter (Gierasch et al., 2000; Ingersoll et al., 2000; Porco et al., 2003). In some SC1 locations there is no distinct MT2 morphology; these may be instances of higher clouds or locally thicker haze overlapping shallower clouds instead. In other locations cellular morphology in CB2 occurs within bright linear MT2 features; these might be regions where outflow from the tops of deep convective cells is spread horizontally, or perhaps regions of convection embedded within larger-scale frontal uplift.

To date, most attention has been focused on the spectacular but sporadic Great Storms seen at 36°S (Dyudina et al., 2007, 2010) and 35°N (Dyudina et al., 2013; Sánchez-Lavega et al., 2011, 2012; García-Melendo et al., 2013; Sayanagi et al., 2013) during the Cassini era. These storms had a considerable effect on the latitude band in which they occurred and on the stratosphere during and after their occurrence. However, the clouds in SC1 are much more common (as is the case on Earth, where small clusters greatly outnumber the largest clusters; see Futyan and Del Genio, 2007) and occur over a wider range of latitudes. They thus may have a greater impact on Saturn's time mean global tropospheric circulation, since they correlate with the apparent rising branches of the mean meridional circulation in the lower troposphere (Ingersoll et al., 2000; Showman et al., 2006; Del Genio et al., 2007; Fletcher et al., 2011a). They also may be diagnostic of conditions at depth that Cassini cannot directly sense. For example, if these clouds are due to water moist convection, the prevalence of SC1 in cyclonic shear regions would suggest that water vapor converges at depth at these latitudes (Del Genio et al., 2009).

We also note that the strong 42°N eastward jet is somewhat weaker ($\sim$10 m s$^{-1}$) in its core and stronger on its cyclonic shear flank in SC1 locations, and to a lesser extent in SC2 and SC3 locations (Fig. 9). We estimate our cloud tracking algorithm to have uncertainties of $\sim$2 m s$^{-1}$ in the mean zonal wind in eastward jet cores and $\sim$1 m s$^{-1}$ elsewhere in the northern hemisphere (Del Genio and Barbara, 2012). The individual zonal wind curves in Figure 9 are based on subsets of the total climatology and thus subject to greater sampling error. On the other hand, the very distinct cellular SC1 features are the easiest to track accurately. The two brightest clusters in MT2 (SC1 and SC2) have wind profiles that are distinct from the two darkest MT2 clusters (SC5 and SC6), suggesting that the differences are real. This is consistent with the observed weakening and spreading of eastward jets with altitude (García-Melendo et al., 2011; Del Genio and Barbara, 2012) and the associated latitudinal temperature profile (Fletcher et al., 2007). It suggests that in CB2 locations with deeper clouds, the cloud-tracked winds are sensing dynamics at higher levels.

Whether the SC1 convective clouds are water- or ammonia-based is unclear. In very favorable thermodynamic conditions (high water abundance, near-saturated relative humidity, dry adiabatic lapse rate), cloud-resolving model simulations produce large, long-lived convective systems reminiscent of some properties of Saturn's Great Storms (Hueso and Sánchez-Lavega, 2004). A water-saturated environment is unlikely to be the norm on Saturn, though (de Graauw et al., 1997) and is more unlikely to occur with a dry adiabatic lapse rate when and where it exists. With a drier environment and/or a lapse rate closer to moist adiabatic, the simulations of Hueso and Sánchez-Lavega (2004) produce transient water convection that is still deep enough to be seen in MT2, and weaker than the Great Storms but still very vigorous by terrestrial standards, and perhaps a better analog for the SC1 clouds. Their simulations of ammonia convection behave similarly for sufficiently high deep ammonia abundance, but only occur for a saturated dry adiabatic environment, which is inconsistent with Cassini microwave retrievals of subsaturated ammonia over much of Saturn (Laraia et al., 2013).

*SC5:* This cluster is the most dramatically different from SC1, being primarily dark in both CB2 and MT2 (Fig. 6) and morphologically consisting primarily of dark linear features or dark ovals (Fig. 8).  Otherwise, though, SC5 is remarkably positively related to SC1 – it occurs about as frequently (6.4%), and at the same latitudes as SC1 (Fig. 9), with an SC1-SC5 latitudinal occurrence correlation of 0.93 (Table 3). Visual inspection of Figures 7 and 8 indicates that many SC5 and SC1 occurrences are close to each other in longitude as well. It is hard to escape the conclusion that these features are causally related to the SC1 convective clouds. Vasavada et al. (2006) found that in Saturn's southern hemisphere, dark vortices observed in CB2 were mostly unaccompanied by similar features in MT2, but the cluster analysis identifies several examples of such features that are obvious at both wavelengths/altitudes (e.g., at 38°N in Fig. 8), with surrounding bright rings, suggesting downwelling in the center and upwelling on the edge, whether or not there is any associated convection.

Since deep convection is accompanied by compensating subsiding motions elsewhere, many SC5 occurrences may represent thinning or clearing of cloud in downwelling regions in response to a nearby deep convective event. Hueso and Sánchez-Lavega (2004) show that simulated subsidence near Saturn convective storms can extend down to 1 bar and below, sufficient to explain the ISS dark features in CB2 and MT2 that surround SC1 convective clouds. Roos-Serote et al. (2000) infer similar downward motion and depletion of water vapor in dark regions adjacent to apparent deep convective clouds on Jupiter. Porco et al. (2005) and Dyudina et al. (2007) show apparent convection possibly injecting vorticity at upper levels and leading to the development of dark spots.  Whether these are due to the same phenomenon or not is unclear. Alternative mechanisms for creating MT2-CB2 dark regions have been proposed that do not rely on cloud thinning or clearing. For example, Baines et al. (2009) find that dark ovals adjacent to a deep convective event absorb over a broad region of the spectrum longward of 0.63 μm. They suggest that lightning-induced chemistry that produces black carbon coating of $NH_3$ and/or $NH_4SH$ ice crystals lofted by the convection can explain the dark regions. On Earth, cluster C5 (Fig. 2), whose properties most closely resemble SC5, captures shallow cumulus clouds that are prevalent, but with fairly low albedo and cloud cover, in the subsiding branch of the Hadley cell, but not spatially connected to individual convective events in the way that SC1 and SC5 occurrences are. We therefore conclude that these Earth and Saturn clusters represent different physical phenomena.

*SC2 and SC3:* These are the two most common clusters on Saturn, together accounting for ~50% of the area in Saturn images (Fig. 6) and occurring fairly often at most latitudes (Fig. 9). They span a broad range of intermediate brightnesses and in general are lower contrast features than the other clusters. Their common occurrence at all latitudes suggests that in some regions they represent a featureless upper troposphere haze that obscures any distinct cloud features below and accounts for Saturn's bland visual appearance relative to Jupiter.

This cannot completely account for the SC2 and SC3 behavior, though. SC2 and SC3 occurrences are highly positively correlated in latitude with each other (0.65, Table 3) but strongly negatively correlated with the occurrence of SC1 (-0.47, -0.73, respectively) and SC5 (-0.36, -0.77). Thus these regions appear unrelated to deep convection. SC2 regions are somewhat brighter than SC3 in CB2, but more dramatically brighter in MT2 (Fig. 6). Morphologically, SC2 and SC3 occur preferentially in the moderately bright, long linear features on the flanks of eastward jets (Fig. 7, Table 4) that account for the familiar chevron cloud patterns on the giant planets. These are the locations of eddy momentum fluxes directed toward the jet core (Del

Genio et al., 2007; Del Genio and Barbara, 2012). Thus these clusters may include clouds of different thicknesses and altitudes produced by baroclinic instability on Saturn. Where this occurs, some fraction of these may be multilayer clouds. The Earth counterparts of these clusters (C2, C3) do not arise from the same process, being related both spatially (Fig. 3) and physically to tropical convective environments. On the other hand, a separate cluster analysis of Earth's extratropics by Williams and Tselioudis (2007) finds clusters with similar radiative properties that they identify with frontal (nimbostratus) and midlevel (altostratus, altocumulus) clouds, respectively, that dominate the midlatitude baroclinically unstable storm tracks. This appears to be a better analog to the Saturn SC2 and SC3 clusters.

*SC4:* These clouds are dark in CB2 but moderately or very bright in MT2 (Figs. 6,8). This suggests that they may be optically thin, high altitude cirrus. On Earth, such clouds (cluster C4) occur climatologically in the same parts of the tropics as deep convection (Fig. 2), but at different times, implying that they form from water vapor detrained from the tops of convective clouds into the environment that later saturates to form ice clouds. On Saturn, it has long been expected that deep convection would inject ammonia or water ice, or vapor that then forms ice, into the upper troposphere, but the upper tropospheric haze has hindered attempts to directly detect such ice spectroscopically, and it is possible that coating of ice by hydrocarbons explains the lack of detection (West et al., 2009). The only positive detections of upper level ice clouds so far have been at times and locations of giant convective storms (Baines et al., 2009; Sromovsky et al., 2013), rather than the thinner cirrus clouds that may be captured by SC4. We cannot rule out the possibility that SC4 is simply capturing upper troposphere haze overlying darker features at lower altitudes. If so, though, we might expect their occurrences to be randomly distributed over the disk (since large-scale latitudinal variations in MT2 albedo are removed by our image processing). Instead, SC4 is most frequent in cyclonic shear regions (Table 4), is moderately positively correlated latitudinally with SC1 and SC5 (0.44, 0.38), and is very negatively correlated with SC2 and SC3 (-0.93, -0.80). Furthermore, inspection of Figure 8 indicates that many SC4 locations are adjacent to SC1 convective cells. This supports the idea that these may indeed be clouds distinct from the haze, produced by detrainment of ammonia and/or water vapor from convection, but occurring where the local environment is more stratified than it is in the convective locations.

*SC6:* These clouds are bright in CB2 but mostly dark or intermediate brightness in MT2 (Fig. 6). Thus they are optically thick clouds with low altitude tops. SC6 latitudinal occurrence is highly positively correlated with SC4 (0.86) and highly negatively correlated with SC2 and SC3 (-0.73,-0.78). It is not quite significantly correlated with SC1 and SC5 (0.23, 0.25), because although SC6 clouds do occur often in cyclonic shear zones they are also common in westward jets (Table 4). SC6 clouds may favor relatively stratified locations, as judged by the variety of MT2 brightnesses in places they occur, which implies that these clouds are unrelated to processes at higher altitude. Their morphology, though, is mostly that of isolated cells (Fig. 8), similar to the appearance of SC1 clouds. Thus these may be shallow cumulus clouds that extend only over short vertical distances and occur when the local overlying troposphere is too dry or has too strong an inversion layer to support deep convection because of large-scale subsidence above. In that sense, SC6 may be most similar to Earth cluster C5.

Shallow cumulus clouds have received virtually no attention in studies of other planets (but see Figures 6 and 14 of West et al., 2016, for a possible exception on Titan), and were ignored on Earth for many years as well. Recently, though, their role in venting moisture from the boundary layer into Earth's lower troposphere and allowing instability to slowly build via large-scale advection has received attention as a contributor to the eventual organization of deep convection on larger spatial scales (Del Genio et al., 2012, 2015). It may be worth considering whether these clouds are another part of the story (in addition to the molecular weight effect of water vapor in a hydrogen atmosphere) of why Great Storms occur so sporadically on Saturn but are so violent and extensive when they do occur (Li and Ingersoll, 2015). In that regard we note that SC6, along with SC4, occurs reasonably often in the broad westward jet region (Fig. 9) where a Saturn Great Storm broke out in 2010 (Sayanagi et al., 2013). Do shallow cumulus "pre-condition" Saturn's atmosphere for later violent events? Are thin cirrus in the westward jet regions the remnants of water injected into the upper troposphere by occasional Great Storms?

*3.3 Temporal variations*

Table 5 shows the RFOs for each cluster for the 4 time periods analyzed. The low-contrast clusters (SC2, SC3) occur less often in 2008 than in 2007, but increase dramatically in 2014. The high contrast clusters with extreme CB2 and/or MT2 brightnesses (SC1, SC4, SC5, SC6), increase in occurrence from 2007 to 2008 but become very rare by 2014.

[Table 5 approximately here.]

Caution should be applied in interpreting these changes, given that the images differ in camera, pixel size, phase angle, and exposure time from one year to another and that the 2014 sample is small (Table 1). Taken at face value, though, they suggest physical changes in Saturn's atmosphere over time. It is possible that the changes are dynamic, e.g., that moist convection increases and then decreases over the course of the mission. A more likely explanation is changes in the optical thickness of Saturn's northern upper tropospheric haze. When Cassini arrived at Saturn in late southern summer, the haze was thicker in the southern hemisphere than the northern hemisphere (West et al., 2009), and image contrasts have been higher in the north than the south since Cassini began (Del Genio et al, 2009). Figure 4 shows that the distribution of relative brightness broadened from 2007 to 2008 but narrowed in 2014.  Fletcher et al. (2016) report warming of the northern hemisphere tropopause layer over the course of the mission but cannot distinguish the effect of increased insolation on a constant haze from the effect of a haze that thickens with time. Our results suggest that the northern haze may indeed be thickening recently, reducing CB2 feature contrasts and explaining the behavior in Figure 4 and Table 5.

## 4. DISCUSSION

*4.1 Saturn's 2010-2011 Great Storm viewed from the clustering perspective*

Use of k-means clustering on Saturn images was motivated primarily by our desire to systematically classify all ISS cloud features, including the less obvious ones that are often ignored.  The interpretations in the previous section are based on the few relationships we have available to guide us, plus the baseline information of how the same procedure identifies cloud types on Earth.  However the interpretations must be considered provisional in the absence of the kinds of detailed information for Saturn that are available to understand the Earth clusters.

One way to test our interpretations is to apply them to an independent, obvious, well-studied feature on Saturn, the Northern Hemisphere Great Storm of 2010-2011 at 33.2°N planetocentric latitude (Fischer et al., 2011; Sánchez-Lavega et al., 2011; Fletcher et al., 2011b; Dyudina et al., 2013; García-Melendo et al., 2013; Sayanagi et al., 2013). The storm was not included in our nominal analysis because of the poorer resolution (107 and 214 km pixel$^{-1}$ for CB2 and MT2, respectively) and viewing geometry (71° phase angle) at the time of the storm. (Had appropriate images been available, it might have defined a seventh independent cluster.)

We applied the same analysis procedure as in Section 2 but adding 1 WAC storm image in each filter from Day 2010-358 (Dec. 24) to our original dataset, with two exceptions. (1) The Great Storm produced extreme high and low MT2 brightnesses (0.090-0.20) outside the 0.115-0.150 range of our other images, implying that it was deeper than the typical SC1 deep convective cells. We thus expanded the range of the highest and lowest MT2 histogram bins. (2) We replicated pixels to map the storm images to the resolution used for the previous dataset. The new k=6 clusters produced in this way are almost identical to those in Figure 6, with correlations between the old and new histograms of 0.997 or higher and very similar RFOs. Thus, for all intents and purposes though not strictly, this is an out-of-sample test, i.e, the classification for the storm images uses clusters that are virtually completely defined by other images.

Figure 10 shows a Great Storm image in CB2 and MT2 with the locations of the 6 clusters superimposed, as in Figures 7 and 8. The associations are very consistent with our previous interpretations for the smaller, more common features. The head of the storm, thought to be the primary location of deep convective updrafts, is primarily classified as SC1. SC2 and SC3, which we argue are characteristic of non-convective regions, are found mostly poleward and equatorward of the storm latitude, with exceptions well east of the storm itself and very few occurrences in the active storm region. SC4 occurs primarily on the fringes of the active storm area, supporting our suggestion that these may be thin cirrus formed from moist air detraining from the tops of the SC1 convection. There are also a few occurrences of SC4 in the trailing wake region where some SC1 cloud features also occur, suggesting that these regions contain localized convection and cirrus. SC5 occurs mostly in the area immediately surrounding the primary convective head. This is consistent with the idea that this cluster captures compensating subsidence in response to the nearby convective upwelling (similar to the Jupiter conclusion of Roos-Serote et al., 2000), though the lightning-produced absorbing aerosol hypothesis of Baines et al. (2009) is another possibility. The SC5 locations surrounding the convective head have a cloud top pressure near 1400 hPa, vs. 400 hPa in the convective head (SC1) region, according to García-Melendo et al. (2013). SC6 is the most randomly distributed cluster, with few occurrences within the head but some everywhere else, including less disturbed parts of the wake region, which plausibly might be expected to contain shallow cumulus clouds.

[Fig. 10 approximately here.]

*4.2 Does MT3 contain independent information?*

It is of interest to know whether processes that determine cloud features in the upper troposphere (sensed by MT2) extend to higher altitudes. The 2007 NAC image sequences used in this paper did not include images in the stronger 889 nm methane band filter (MT3), which senses the tropopause/lower stratosphere, due to Cassini solid state recorder limitations. However the 2008 WAC images contain all three (CB2, MT2, MT3) filters. The small MT3 sample size makes it non-ideal for cluster analysis, but some information can be gathered in a different way.

To get a sense of whether MT3 behaves much differently than MT2, Figure 11 shows a false color image obtained by combining images in the three filters for Day 2008-128 and assigning red, green, and blue to CB2, MT2, and MT3, respectively. Features unique to MT3 and not also captured by MT2 are colored either blue (bright in MT3 but dark in the other filters), yellow (dark in MT3 but bright in the other filters), violet (bright in MT3 and CB2 but dark in MT2), or green (bright in MT2 but dark in MT3 and CB2). Examples of each can be seen in Figure 11, e.g., in the wavelike feature on the northern flank of the 42°N eastward jet, the chevron features south of the 34°N westward jet, and the bright collar surrounding the dark vortex at 55-60°N. This suggests that several new clusters might be defined by a 3-filter analysis were sufficient images available, but that the RFOs of such clusters would be small relative to those we have already defined.

[Fig. 11 approximately here.]

To further explore the similarities between MT2 and MT3 features, we started with the original 6 CB2-MT2 clusters (Fig. 6). For each cluster geographic location in the image in Figure 11, we then created a frequency histogram of MT3-CB2 brightness values (after processing the MT3 image in the same way as described in Section 2 for CB2 and MT2 images). The result (Fig. 12) shows that the properties of these histograms are qualitatively similar to those of the MT2-CB2 clusters in Figure 6, i.e., clusters that are preferentially bright in MT2 (SC1, SC2, SC4) also tend to be bright in MT3, clusters of intermediate MT2 brightness (SC3) are also intermediate in MT3, and clusters dark in MT2 (SC5, SC6) are also dark in MT3. Figure 13 shows the complementary set of histograms for MT3 vs. MT2. In general the points lie along the diagonal defined by similar MT3 and MT2 brightness. The largest off-diagonal contributions appear to be in SC2, SC3, and SC4. These may be the best places to search for behavior unique to MT3 that would isolate processes restricted to the near-tropopause region.

[Fig. 12 approximately here.]

[Fig. 13 approximately here.]

*4.3 Relationship to the Thompson (1990) clusters*

Thompson's (1990) k-means clustering analysis of Voyager 2 images is not directly comparable to ours for several reasons. It was for Jupiter rather than Saturn, it relied strictly on visible wavelength images, and it was performed on individual pixels over 4 filters to derive photometric information about color/albedo, rather than the meteorological emphasis of our Saturn study over regions encompassing many pixels.  Given that emphasis, Thompson chose to identify a large number of clusters (25) rather than our smaller dynamics-motivated set, although 98% of all Jupiter pixels are in his first 10 clusters. Nonetheless, examination of Thompson's clusters suggests some analogs between the cloud types he identified and ours.

Thompson's cluster 7 (5% of his pixels) is the brightest unit that is bright across all 4 filters (including a weak methane band at 621 nm). It includes small equatorial plumes and bright features in the South Equatorial Belt, a cyclonic shear region in which deep convective clouds are seen in Cassini Jupiter flyby images (Porco et al., 2003).  This appears to be the best analog for our SC1 cluster. SC2 may have some things in common with Thompson's units 5-6, which occur in zones (anti-cyclonic shear regions) and are not as bright as his unit 7.  However his units 1-3, which include 75% of all Jupiter pixels and represent the "average" Jupiter, may contain areas most similar to our SC2 and especially SC3, the most neutral of all our clusters, which together make up ~50% of our Saturn area. Thompson's brightest units in the methane band filter relative to the green filter are 14, 17-20, and 23, but it is not clear whether these are related to our SC4; 14 and 17 are reported as being in the South Equatorial Belt not far from the Great Red Spot, with 17 having "additional thin cloud cover."  Our SC5 may be most analogous to Thompson's units 8 and 9, which are dark in all filters and common in belts (cyclonic shear regions).  Finally, our SC6 may be most similar to Thompson's units 4-6, which are dark in the methane band but bright in the green and thus seem to capture low altitude clouds.

## 5. CONCLUSION

We have demonstrated the potential utility of k-means clustering to organize and exploit *all* the information content in images of giant planet atmospheres.  Several benefits may flow from the type of classification we have performed. As Thompson (1990) first imagined, clustering creates a database of spectrally similar features that may be used as a guide to the planet, allowing spatially separated but physically similar regions to be selected for analysis together. This would allow the user to derive cloud structure models in many filters that are simultaneously more generally representative of the planet as a whole but with sufficient detail to identify distinct "units" worthy of separate analysis. Since our approach focuses on small-scale features rather than large-scale albedo contrasts, its use would be as the starting point for a quantitative analysis of the vertical cloud structure of these small features. Each data subset, based on many occurrences of a given cloud type, would when aggregated have higher signal-to-noise ratio than a single image scene and also allow real variability within a given cloud spectral/structural type to be characterized. For example, the search for evidence of ammonia or water ice clouds in Saturn's upper troposphere might be facilitated by gathering observations that are collocated with a particular ISS cloud type (in this case, SC4) and combining them to better detect a faint spectral signal. A similar approach might be applied to systematically relate gaseous composition anomalies to deep convective clouds (SC1) or to simply estimate the top pressures of these clouds relative to the larger-scale cloud deck. Likewise, a systematic near-infrared analysis of SC5 occurrences surrounding bright SC1 clouds, in dark ovals in the general vicinity of SC1 clouds, and in association with more linear features might be able to distinguish examples due to downwelling holes in the upper level clouds and haze vs. absorbing contaminants on the surfaces of $NH_3$ and/or $NH_4SH$ ice crystals formed by lightning-induced chemistry as suggested by Baines et al. (2009).

For the specific clustering approach we use, an additional benefit is how it facilitates comparative planetology. We have seen that although there are many differences between Saturn and Earth, they actually share many similar phenomena as indicated by the clusters that emerge for each planet.  For Saturn, our analysis suggests that on the large scale, the atmosphere can be divided into three dynamically distinct types of latitudinal bands: (1) Cyclonic shear regions poleward of the eastward jets, which are populated by cloud types indicative of deep convectively disturbed conditions (SC1, SC5, SC4); (2) Generally suppressed regions near and on either side of the broad westward jets, in which shallow convective clouds (SC6) are common, implying dry conditions at high altitude, except for rare outbreaks of Great Storms; (3) Baroclinically unstable regions near the cores of the eastward jets and encompassing the anti-cyclonic shear regions on their equatorward flanks (SC2, SC3).  These are roughly analogous to Earth's tropics, subtropics, and midlatitudes, respectively.  This may be a more useful classification than the traditional belt-zone partitioning, which in any case does not work as well for Saturn as for Jupiter because of the thick upper troposphere haze. The latitudinal structure of the haze is better related to processes occurring near and above the tropopause (Roman et al., 2013) than to the dynamics of the troposphere. Figure 14, which shows the appearance of the sample images in Figures 7-8 before large-scale albedo variations are removed, illustrates that the well-known large-scale albedo bands in MT2 are not well correlated with those in CB2 and that neither correlates well with our dynamically-defined bands, which are based on small-scale feature contrasts and morphology.

[Fig. 14 approximately here]

Water vapor has been detected at subsaturated levels in Saturn's troposphere (de Grauw et al., 1997) but has not been mapped at depth. Our classification anticipates what this distribution may look like by analogy to Earth's troposphere, if the SC1 deep convective clouds we see are driven by water condensation. If the SC1 clouds are water-based convection, we might expect that near and above the base of the water condensation level, the atmosphere is closest to saturated in the cyclonic shear latitudes, where moisture converges, deep convection is common, and the mean meridional circulation rises, while the westward jet regions are only moderately moist at depth and considerably drier at higher altitudes, where the mean circulation subsides. Such latitudinal variations might be detected on Jupiter by the upcoming Juno mission. We note that Jupiter may be an even better target for clustering analysis than Saturn, since the absence of a thick upper tropospheric haze there allows for more sensitive altitude discrimination using continuum and methane band filters than is possible on Saturn.

We finally note that the fact that Thompson (1990) published his vision of the use of cluster analysis on giant planet images in a high-performance computing journal suggests the considerable computational effort required at that time to perform such an analysis on only 44 images. Twenty-five years later, we have easily analyzed 200 images on a typical desktop computer, limited by the availability of high resolution northern hemisphere CB2-MT2 pairs rather than by computational constraints. Rossow et al. (2005) have applied their cluster analysis for Earth to almost 30 years of data. We suggest that the time is right for this type of objective analysis of planetary atmosphere images to become routine for any spacecraft mission.

**ACKNOWLEDGEMENTS.** This research was supported by Cassini Project funding of the Imaging Science Subsystem team. We thank Robert West for helpful suggestions about interpretations of the clusters. We also thank two reviewers for constructive comments. The clusters obtained from the research described in this paper are available from data.giss.nasa.gov/cassini_clusters. The k-means clustering software is available from isccp.giss.nasa.gov/tcluster.html.

Table 1. Details of the images used in the cluster analysis.

| Date | Camera | Filter | Number of images | Image scale (km pixel$^{-1}$) | Phase angle | Exposure duration (ms) | Gain state (e/DN) |
|---|---|---|---|---|---|---|---|
| 2007-310 | NAC | CB2 | 72 | 17 | 38° | 2600 | 29 |
| (6 Nov.) | NAC | MT2 | 72 | 35 | 38° | 3800 | 29 |
| 2008-108 | WAC | CB2 | 14 | 82 | 38° | 260 | 29 |
| (17 April) | WAC | MT2 | 14 | 82 | 38° | 1500 | 29 |
| 2008-128 | WAC | CB2 | 14 | 76 | 46° | 260 | 29 |
| (7 May) | WAC | MT2 | 14 | 76 | 46° | 1500 | 29 |
| 2014-259 | WAC | CB2 | 2 | 89 | 20° | 380 | 29 |
| (16 Sept.) | WAC | MT2 | 2 | 89 | 20° | 1800 | 29 |

Table 2. Correlations between cluster histogram patterns for cluster analyses using N = 5 and 6 clusters.

k = 5 clusters

|     | SC1 | SC2 | SC3 | SC4 | SC5 |
| --- | --- | --- | --- | --- | --- |
| SC1 | X | | | | |
| SC2 | -0.19 | X | | | |
| SC3 | -0.23 | -0.34 | X | | |
| SC4 | -0.15 | -0.28 | 0.25 | X | |
| SC5 | 0.16 | -0.40 | -0.66 | -0.12 | X |

k = 6 clusters

|     | SC1 | SC2 | SC3 | SC4 | SC5 | SC6 |
| --- | --- | --- | --- | --- | --- | --- |
| SC1 | X | | | | | |
| SC2 | 0.16 | X | | | | |
| SC3 | -0.36 | -0.06 | X | | | |
| SC4 | -0.20 | -0.31 | -0.26 | X | | |
| SC5 | -0.11 | -0.33 | -0.05 | 0.28 | X | |
| SC6 | 0.27 | -0.29 | -0.34 | -0.56 | -0.16 | X |

Table 3. Correlation coefficients between the number of occurrences of each Saturn cluster at a given latitude and the number of occurrences of all other clusters. A correlation of 0.30(0.39) is significant at the 95(99)% level.

| | SC1 | SC2 | SC3 | SC4 | SC5 | SC6 |
|------|------|------|------|------|------|-----|
| SC1 | X | | | | | |
| SC2 | -0.47 | X | | | | |
| SC3 | -0.73 | 0.65 | X | | | |
| SC4 | 0.44 | -0.93 | -0.80 | X | | |
| SC5 | 0.93 | -0.36 | -0.77 | 0.38 | X | |
| SC6 | 0.23 | -0.73 | -0.78 | 0.86 | 0.25 | X |

Table 4. Ratio of number of occurrences per degree latitude of each cluster for each latitude band type to the number of occurrences per degree for the entire analysis domain.

| | SC1 | SC2 | SC3 | SC4 | SC5 | SC6 |
|---|---|---|---|---|---|---|
| Cyclonic shear | 1.43 | 0.79 | 0.69 | 1.33 | 1.46 | 1.18 |
| Eastward jet + anti-cyclonic shear | 0.68 | 1.30 | 1.39 | 0.48 | 0.67 | 0.67 |
| Westward jet | 0.68 | 1.02 | 1.08 | 1.02 | 0.64 | 1.06 |

Table 5. Relative frequency of occurrence (RFO) of each Saturn cluster for the 4 years represented in the analyzed data.

| Cluster # | Relative Frequency of Occurrence (%) | | | |
|:---:|:---:|:---:|:---:|:---:|
| | 2007-310 | 2008-108 | 2008-128 | 2014-259 |
| 1 | 3.4 | 7.3 | 8.2 | 0.8 |
| 2 | 25.3 | 19.9 | 19.3 | 26.5 |
| 3 | 41.5 | 26.4 | 22.2 | 65.6 |
| 4 | 12.8 | 19.3 | 18.9 | 2.0 |
| 5 | 4.4 | 6.7 | 8.6 | 1.8 |
| 6 | 12.6 | 20.5 | 22.8 | 3.3 |

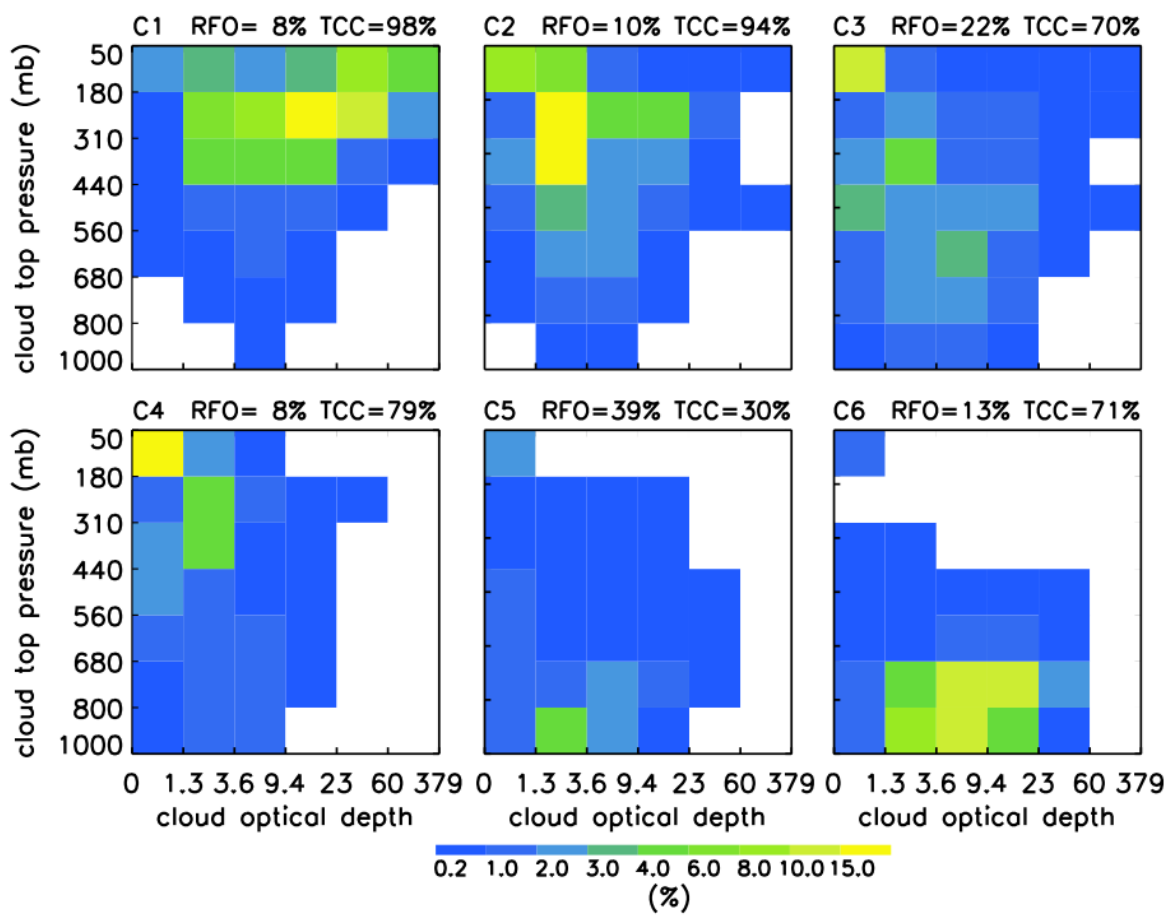

Figure 1. Weather states obtained from k-means clustering of Earth geostationary visible and window infrared satellite retrievals of cloud top pressure and optical depth for the 15°N-15°S latitude band. The numbers above each histogram indicate the relative frequency of occurrence (RFO) and total cloud cover (TCC) for each cluster. (Reproduced with permission from Figure 1 of Chen and Del Genio, 2009, Springer).

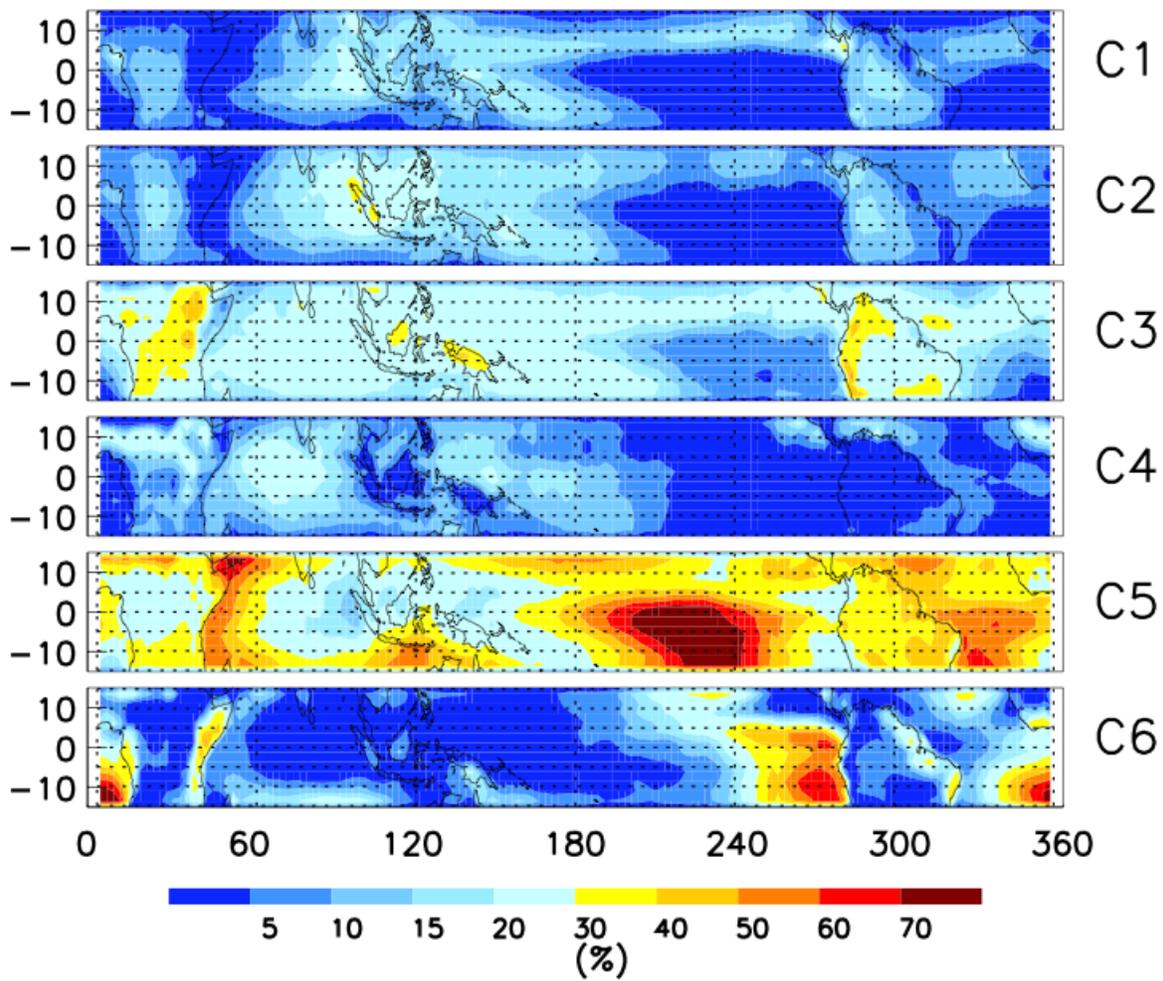

Figure 2. Geographic distribution of the relative frequency of occurrence of the 6 Earth clusters shown in Figure 1. (Reproduced with permission from Figure 2 of Chen and Del Genio, 2009, Springer)

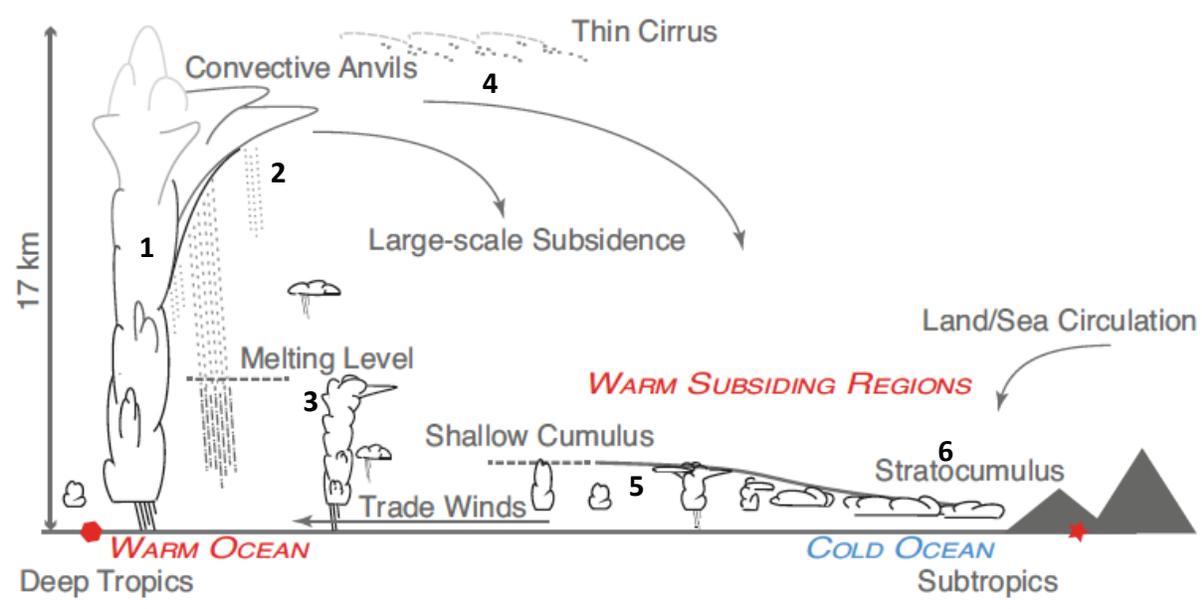

Figure 3. Schematic cross-section across Earth's tropics and subtropics showing the dominant cloud types in the rising and subsiding branches of the Hadley circulation. The numbers indicate the clusters in Figure 1 that correspond to each cloud type. (Adapted with permission from Figure 7.4(c) of Boucher et al., 2013.)

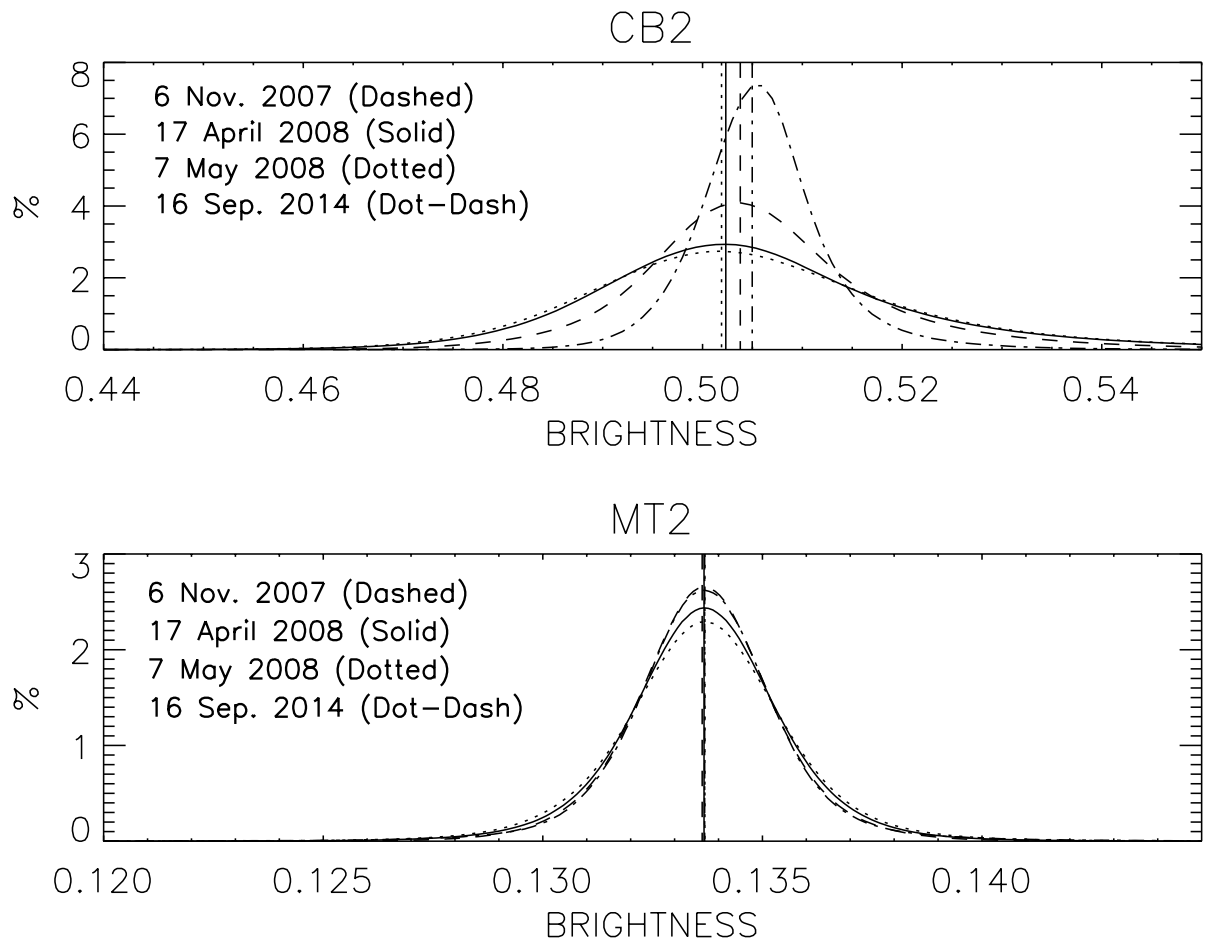

Figure 4. Distributions of brightness values for each image set. The vertical lines represent the mode values for each distribution. Top: CB2. Bottom: MT2.

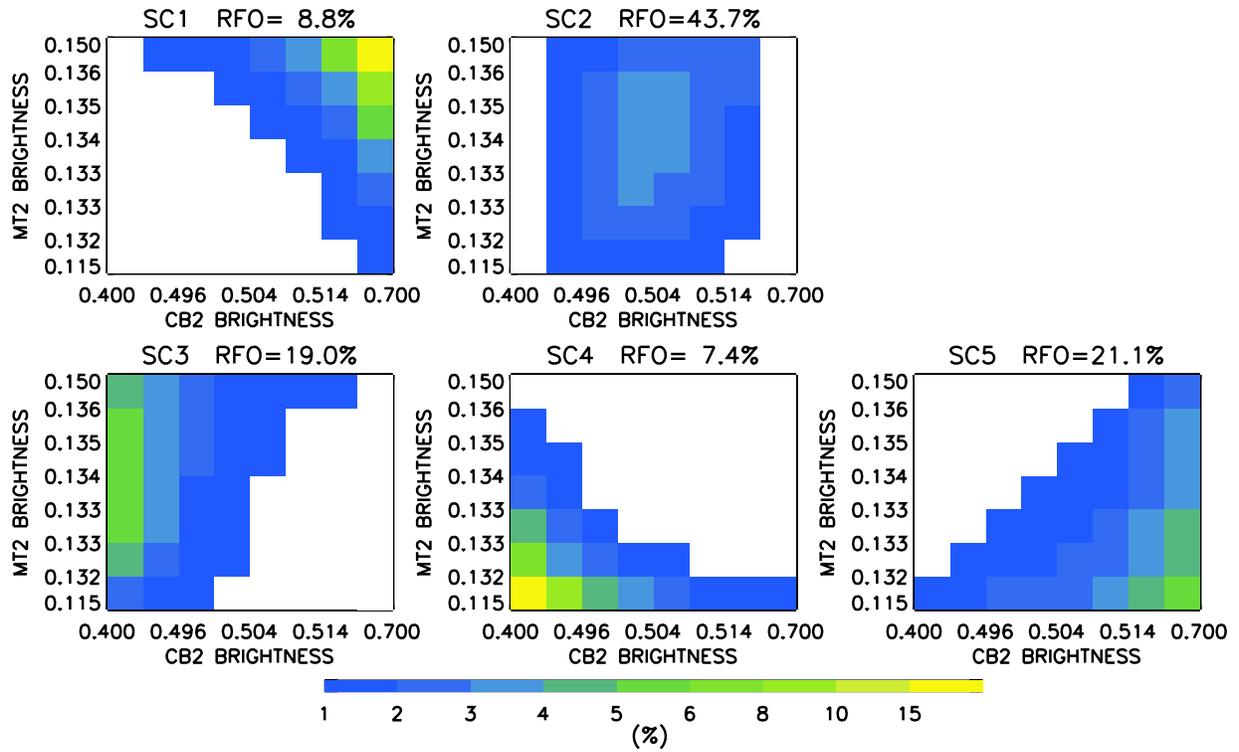

Figure 5. As in Figure 1, but for clusters obtained from analysis of Cassini ISS MT2 vs. CB2 I/F histograms when k = 5 is assumed.

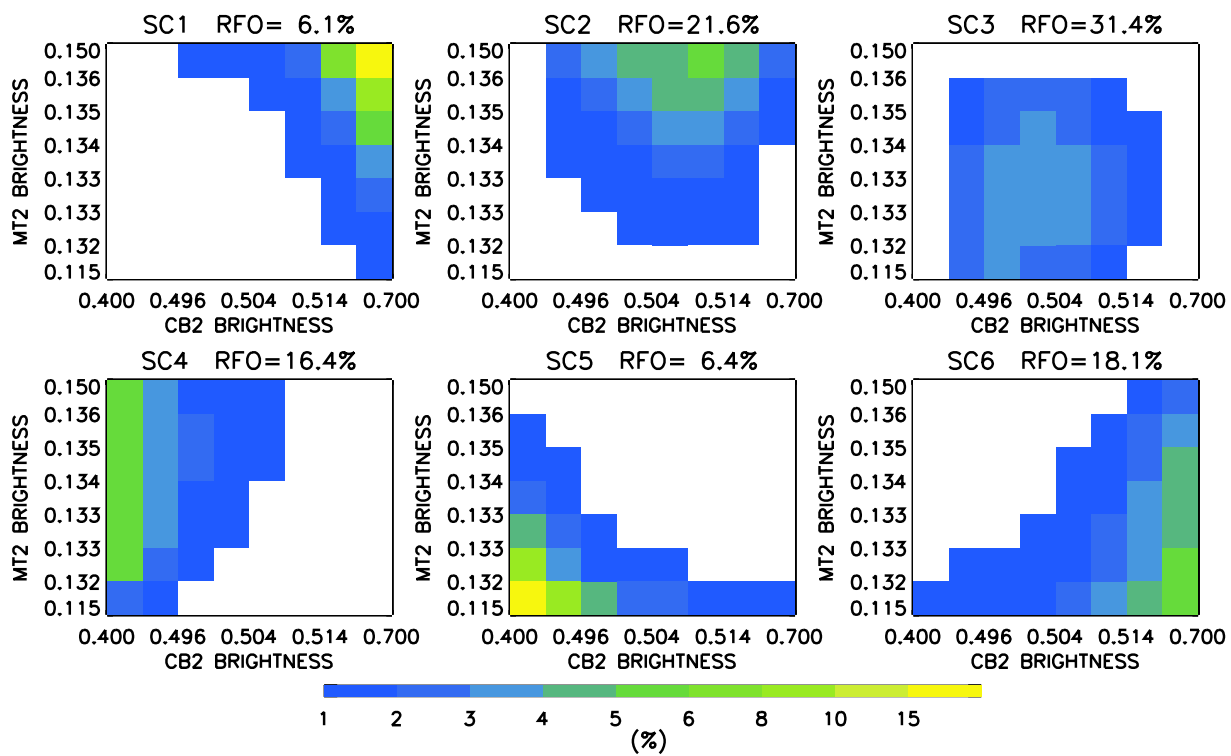

Figure 6. As in Figure 5, but for the assumption k = 6.

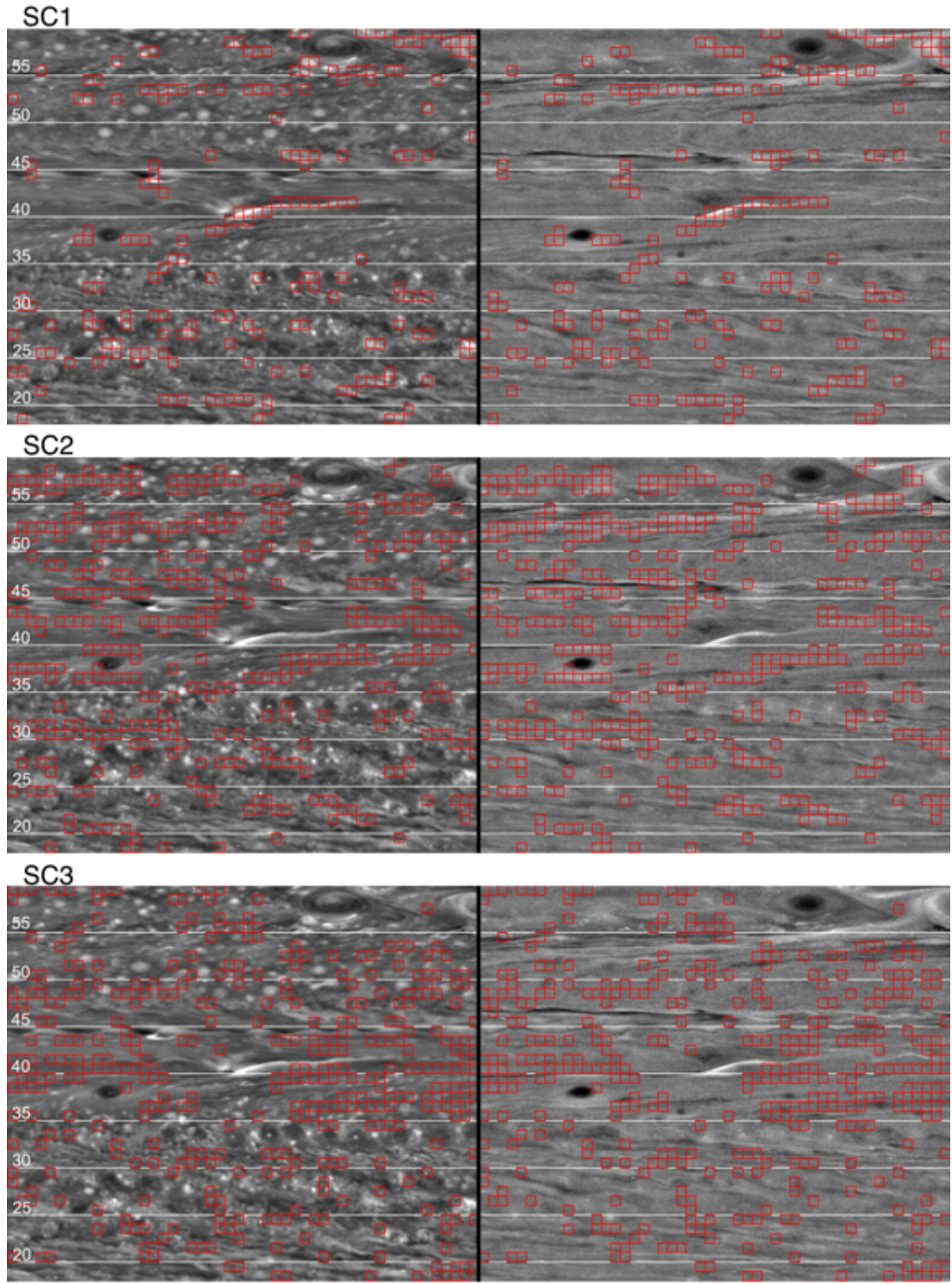

Figure 7. Sample Saturn CB2 (left) and MT2 (right) processed WAC images of the 18°N-60°N region, with occurrences of clusters (top) SC1, (middle) SC2, and (bottom) SC3 identified by the overlaid red boxes.

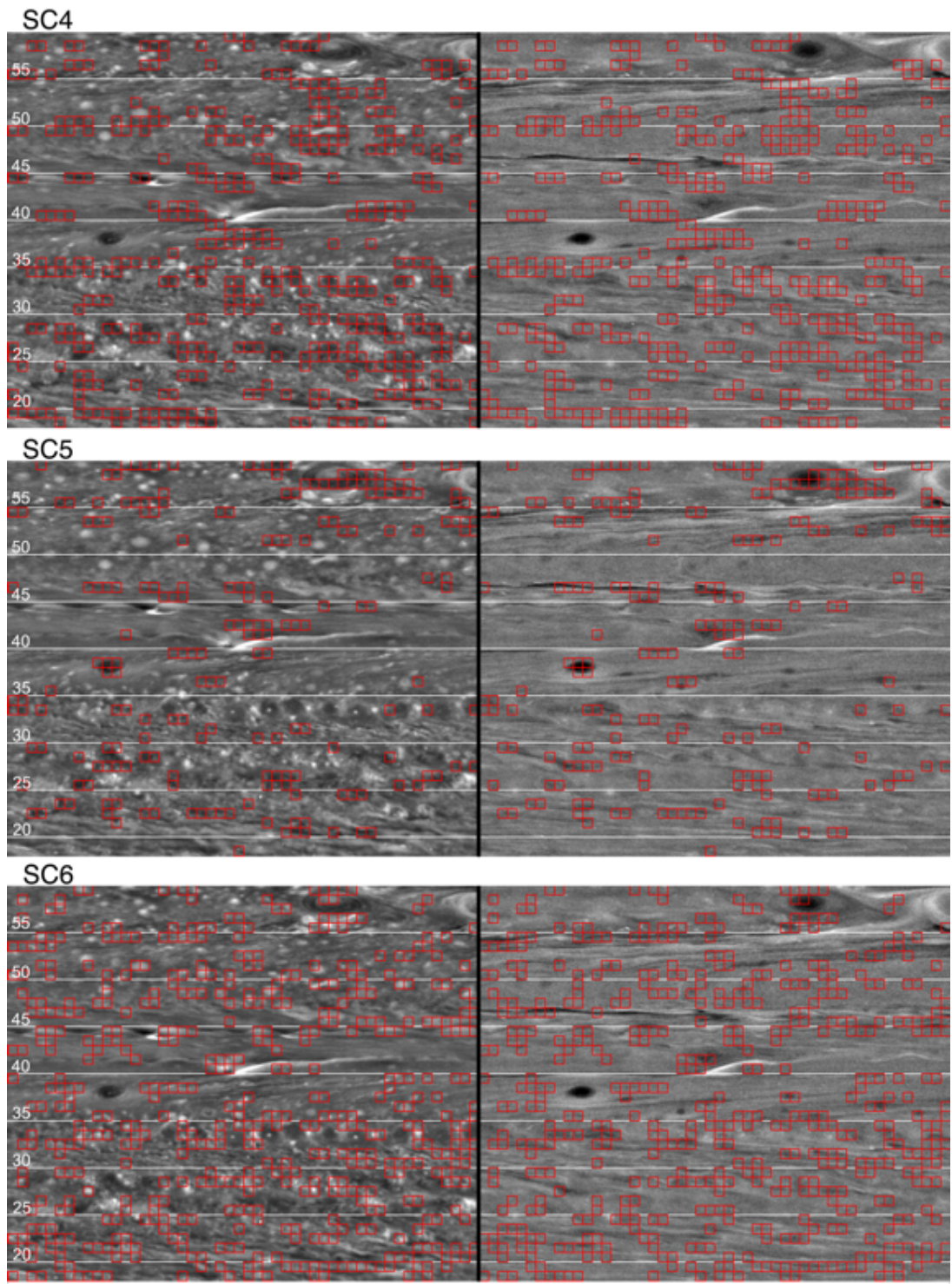

Figure 8. As in Figure 7 but for Saturn clusters (top) SC4, (middle) SC5, (bottom) SC6.

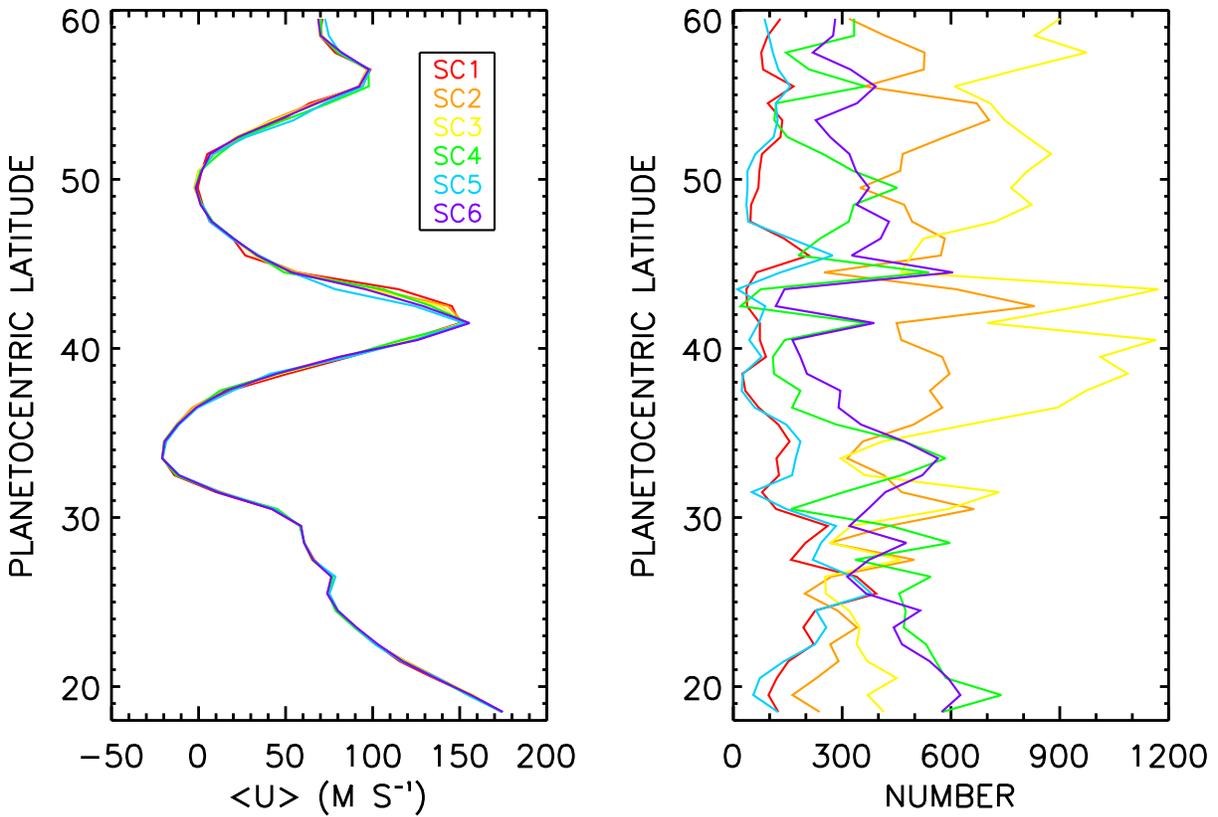

Figure 9. Left: Zonal mean CB2 cloud tracked wind profiles for tracking target boxes associated with each of the 6 Saturn clusters. Right: Number of occurrences of each cluster as a function of latitude. (Note that not every occurrence of a cluster has a successful wind estimate.)

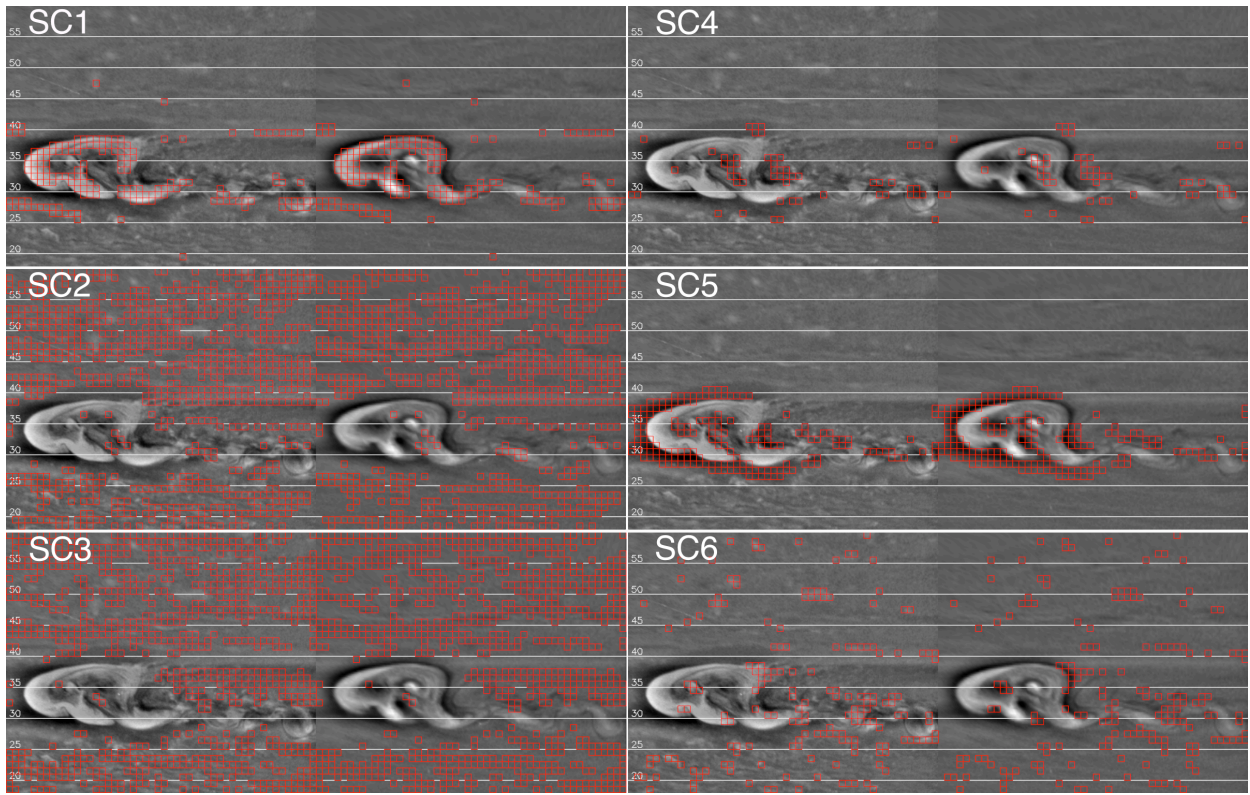

Figure 10. As in Figures 7 and 8, but for the Saturn Northern Hemisphere great storm of 2010-2011.

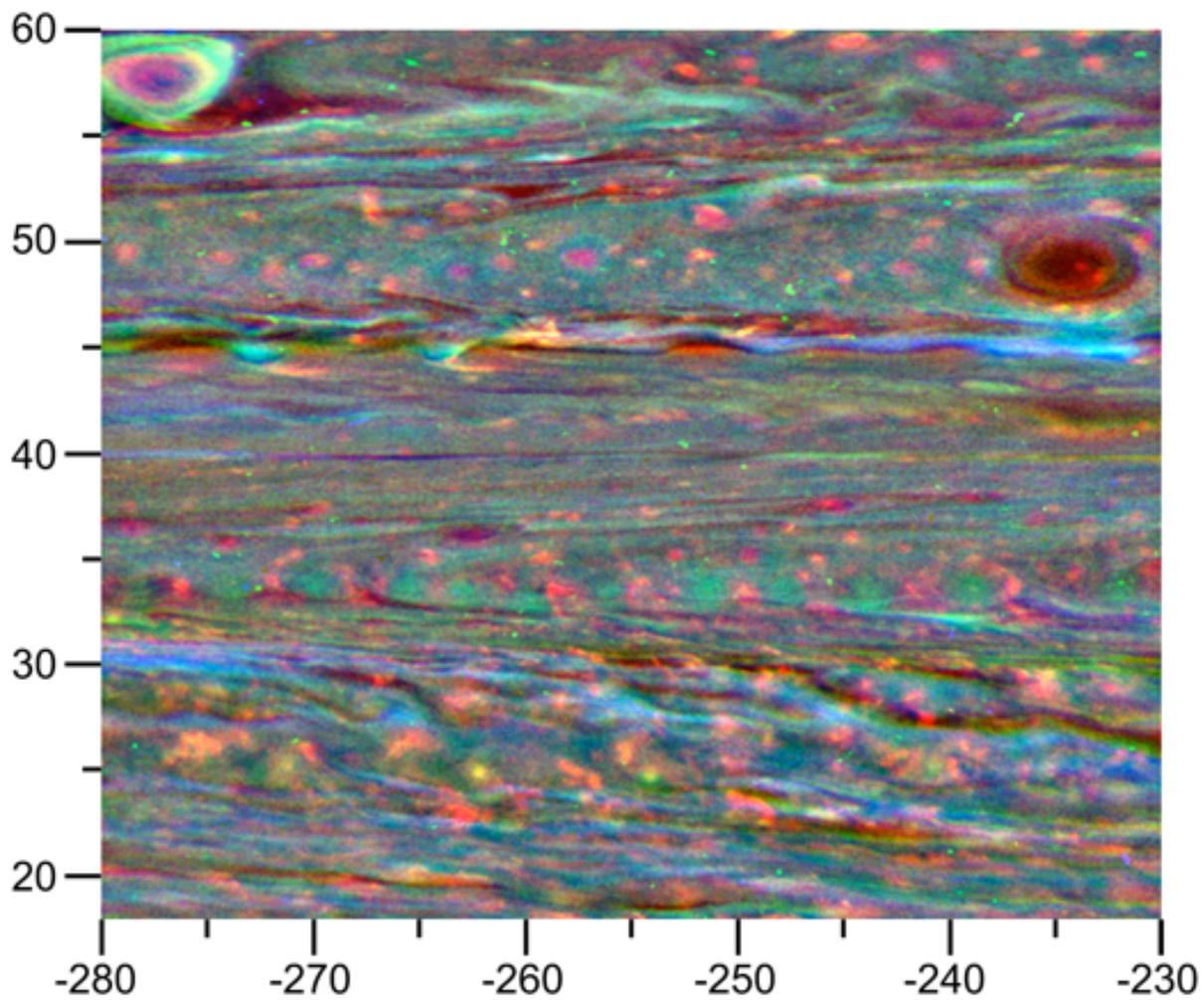

Figure 11. CB2 (red) – MT2 (green) – MT3 (blue) false color WAC image acquired on Day 2008-128.

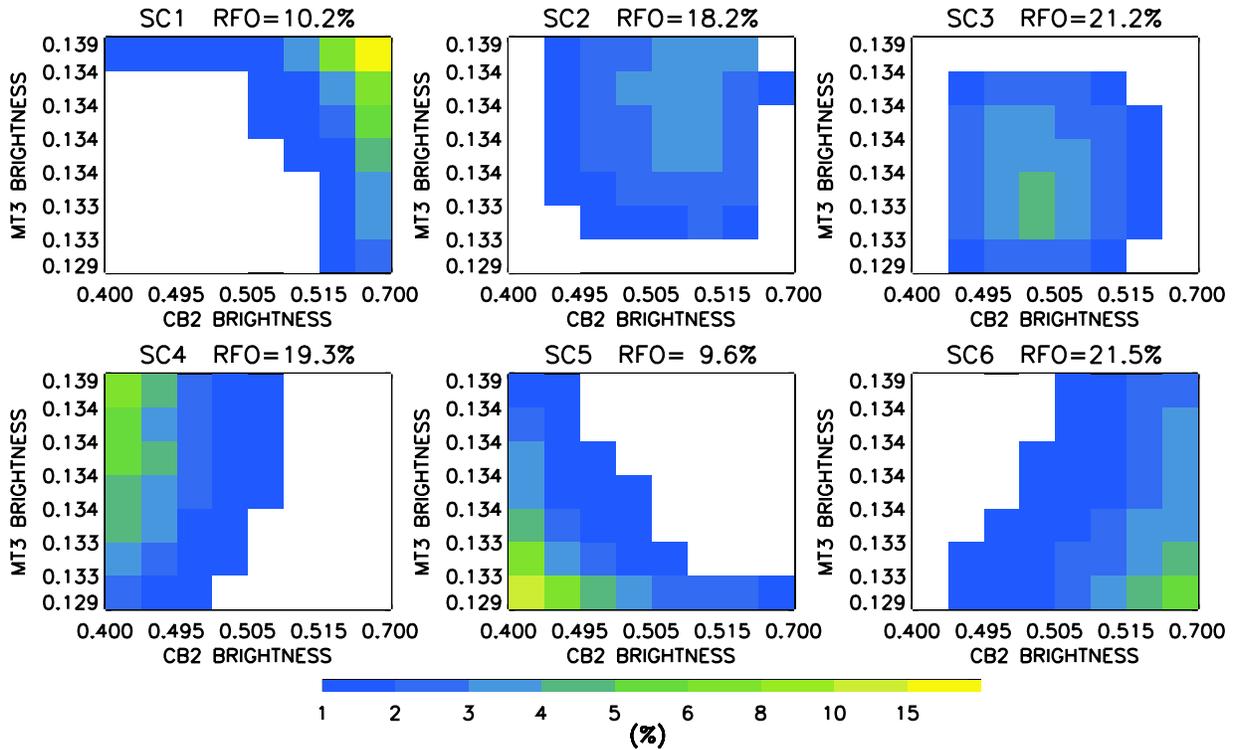

Figure 12. MT3-CB2 I/F frequency histograms obtained from the images used in Figure 11. The histograms were constructed by classifying the regions in the figure using the MT2-CB2 clusters defined in Figure 6 and then mapping the coincident MT3 I/F values into the appropriate cluster. Note that the RFOs of these clusters differ from those in Figure 6 because they are based only on a single image rather than the full dataset.

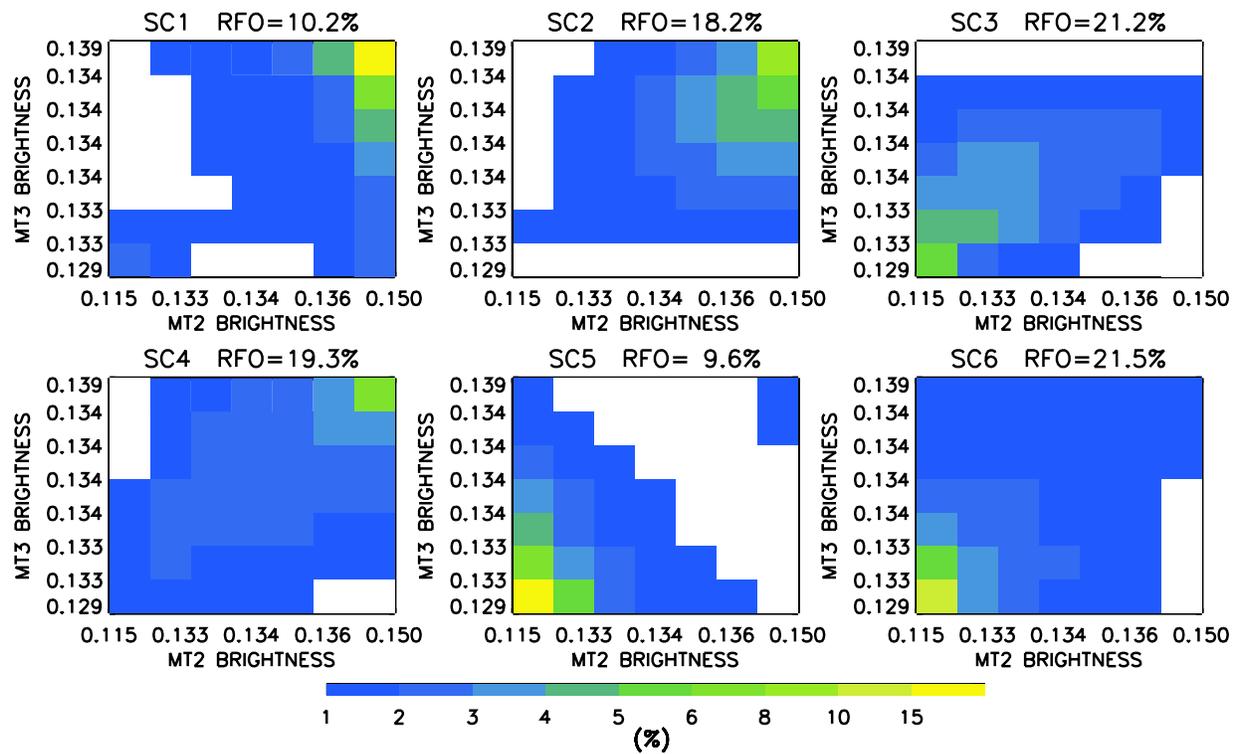

Figure 13. As in Figure 12 but for MT3 vs. MT2.

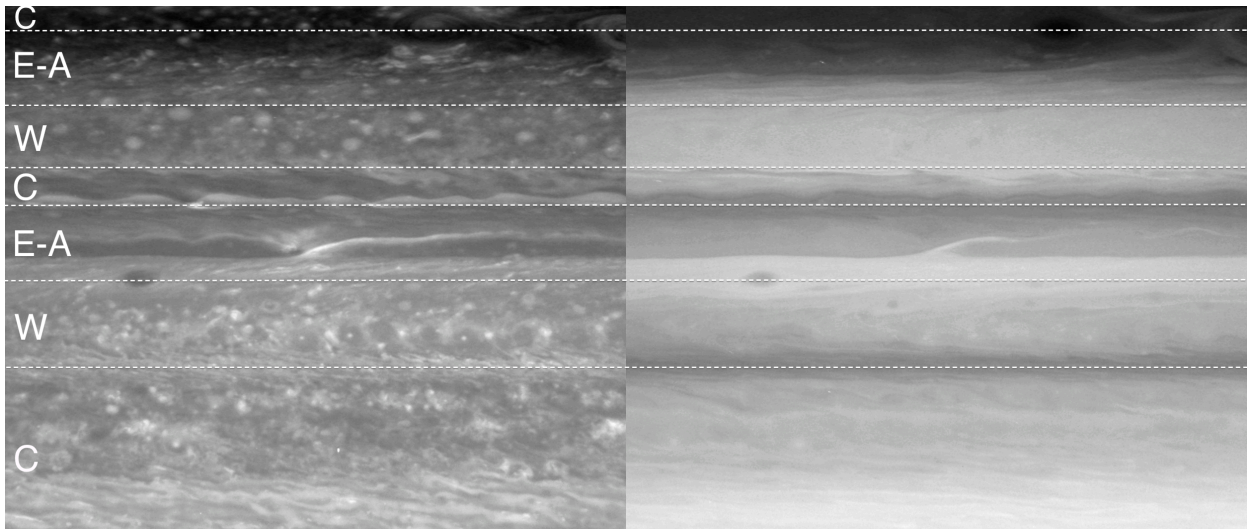

Figure 14. Original versions of the (left) CB2 and (right) MT2 images shown in Figures 7 and 8, with only the Minnaert function processing to reduce illumination gradient effects and no removal of large-scale albedo contrasts. The horizontal dotted lines indicate the boundaries of our suggested dynamically-based band classification (C = cyclonic shear; E-A = eastward jet + anticyclonic shear; W = westward jet).